\documentclass[eqsecnum,twocolumn,amsmath,showpacs,amsfonts,aps,prc,%
floatfix]{revtex4}

\usepackage{graphicx}
\usepackage{bm}

\begin{document}


\title{Dissipative hydrodynamics for viscous relativistic fluids}

\author{Ulrich Heinz}
\email[E-mail:]{heinz@mps.ohio-state.edu} 
\author{Huichao Song}
\email[E-mail:]{song@mps.ohio-state.edu} 
\affiliation{Department of Physics, The Ohio State University, Columbus, 
OH 43210, USA}
\author{A. K. Chaudhuri}
\email[E-mail:]{akc@veccal.ernet.in}
\affiliation{Variable Energy Cyclotron Centre, 1/AF, Bidhan Nagar, 
Kolkata 700~064, India}

\begin{abstract}
Explicit equations are given for describing the space-time evolution
of non-ideal (viscous) relativistic fluids undergoing boost-invariant
longitudinal and arbitrary transverse expansion. The equations are
derived from the second-order Israel-Stewart approach which ensures 
causal evolution. Both azimuthally symmetric (1+1)-dimensional and
non-symmetric (2+1)-dimensional transverse expansion are discussed.
The latter provides the formal basis for the hydrodynamic computation 
of elliptic flow in relativistic heavy-ion collisions including 
dissipative effects.
\end{abstract}


\date{\today}  

\maketitle

\section{Introduction}
\label{sec1}
%
Ideal fluid dynamics has been used successfully to predict the 
collective flow patterns in Au+Au collisions at the Relativistic 
Heavy-Ion Collider RHIC (for a review see \cite{QGP3}). The ideal
fluid description works well in almost central Au+Au collisions near 
midrapidity at top RHIC energy, but gradually breaks down in more 
peripheral collisions, at forward rapidity, or at lower collision
energies \cite{Heinz:2004ar}, indicating the onset of dissipative effects.
To describe such deviations from ideal fluid dynamics quantitatively, 
and to use the experimental data for the extraction of values or 
phenomenological limits for the transport coefficients of 
the hot and dense matter created during the collision, requires the 
numerical implementation of {\em dissipative} relativistic fluid dynamics. 
Although a formulation of such a theory which avoids the longstanding 
problems of acausal signal propagation and other instabilities associated
with the original relativistic fluid equations given by Eckart \cite{Eckart}
and Landau and Lifshitz \cite{LL63}, has been known for almost 30 years 
\cite{IS79}, significant progress towards its numerical implementation 
has only been made very recently \cite{Muronga:2001zk,Teaney:2004qa,MR04,CH05}.
At this point, we are only at the very beginning of a program that
will eventually apply viscous relativistic fluid dynamics to heavy-ion 
collision data. Existing numerical implementations are (1+1)-dimensional 
and can only describe cylindrically symmetric transverse expansion with 
boost-invariant longitudinal dynamics \cite{MR04,CH05}. As we will show
here, even the (1+1)-dimensional case still presents some open formal
issues which we address in the present paper. The numerical codes are 
still in the process of being tested and will not be discussed here. 

The paper is organized as follows: In Section~\ref{sec2} we shortly
review relativistic ideal fluid dynamics and the conditions for its
applicability. While most of this is standard textbook material, it
helps to establish notation and to better appreciate the differences
in the non-ideal case. In Section~\ref{sec3} we discuss the non-ideal
fluid decomposition and introduce the dissipative flows (bulk and shear
viscous pressure, heat conduction) and how they manifest themselves
in the baryon current and energy-momentum tensor in the Eckart and 
Landau frames. Section~\ref{sec4} deals with the derivation of 
equations to determine the evolution of these dissipative flows. We 
follow the treatment of Israel and Stewart \cite{IS79} and discuss
both the (acausal) first-order and (causal) second-order theories
(this nomenclature will be explained in Sec.~\ref{sec4}). While most
of the material up to this point can already be found elsewhere
\cite{IS79,Muronga:2001zk} (see also the beautiful lecture notes by
Rischke in \cite{Rischke}), it is needed here for a selfcontained
presentation and for a critical discussion of some systematic
expansion issues which we point out in Sec.~\ref{sec4b} and which
are of practical relevance. Section~\ref{sec5} contains the main
results of this paper (with many technical details deferred to the
Appendix), namely complete sets of causal equations of motion for
the dissipative transverse hydrodynamic expansion of systems undergoing
boost-invariant longitudinal flow. The discussion of the azimuthally
symmetric (1+1)-dimensional case in Sec.~\ref{sec5a} improves on the
presentation given in the recent work by Muronga and Rischke \cite{MR04},
while the equations for the non-symmetric (2+1)-dimensional case
in Sec.~\ref{sec5b} are original and have, to our knowledge, not been
presented before. The concluding Section~\ref{sec6} summarizes our 
results and gives some further discussion.

\section{Ideal fluid dynamics}
\label{sec2}
%
Before explaining the structure of the equations for causal dissipative
relativistic fluid dynamics, let us quickly review the case of ideal
fluid dynamics. Any fluid dynamical approach starts from the
conservation laws for the conserved charges and for energy-momentum,
\begin{eqnarray}
\label{eq1}
\partial_\mu N^\mu_i &=& 0, \quad i=1,\dots,k,
\\
\label{eq2}
\partial_\mu T^{\mu\nu} &=& 0.
\end{eqnarray}
For simplicity we will restrict ourselves to $k{\,=\,}1$ (say, $N_\mu=$ 
net baryon number current) and drop the index $i$ in (\ref{eq1}).
It must also ensure the second law of thermodynamics
\begin{equation}
\label{eq3}
\partial_\mu S^\mu \geq 0,
\end{equation}
where $S^\mu$ is the entropy current. Ideal fluid dynamics follows from
these equations under the assumption of local thermal equilibrium, i.e.
if the microscopic collision time scale is very much shorter than any 
macroscopic evolution time scale such that the underlying phase-space 
distribution $f(x,p)$ relaxes essentially instantaneously to a local
equilibrium form (upper signs for fermions, lower signs for bosons)
\begin{equation}
\label{eq4}
f_{\rm eq}(x,p) = \frac{1}{e^{[p{\cdot}u(x)+\mu(x)]/T(x)}\pm 1},
\end{equation}
where $u^\mu(x)$ is the local fluid velocity at point $x$, $\mu(x)$ is 
the local chemical potential associated with the conserved charge 
$N$ (it enters with opposite sign in the distribution $\bar f$ for
antiparticles), and $T(x)$ is the local temperature. Plugging this into 
the kinetic theory definitions 
\begin{eqnarray}
\label{eq5}
N^\mu(x) &=& \int \frac{d^3p}{E} p^\mu [f(x,p)-\bar f(x,p)],
\\
\label{eq6}
T^{\mu\nu}(x) &=& \int \frac{d^3p}{E} p^\mu p^\nu [f(x,p)+\bar f(x,p)],
\\
\label{eq7}
S^\mu(x) &=& - \int \frac{d^3p}{E} p^\mu 
            \Bigl[f(x,p)\ln f(x,p)
\\
\nonumber
&&\quad\pm\bigl(1{\mp}f(x,p)\bigr)\ln\bigl(1{\mp}f(x,p)\bigr) + 
                 (f\leftrightarrow\bar f)\Bigr],
\end{eqnarray}  
leads to the ideal fluid decompositions
\begin{eqnarray}
\label{eq8}
&&N_{\rm eq}^\mu = n\, u^\mu,
\\
\label{eq9}
&&T_{\rm eq}^{\mu\nu} = e\,u^\mu u^\nu - p\, \Delta^{\mu\nu} \ 
(\mbox{with}\ \Delta^{\mu\nu}{=}g^{\mu\nu}{-}u^\mu u^\nu),\qquad
\\
\label{eq10}
&&S_{\rm eq}^\mu = s\, u^\mu,
\end{eqnarray}  
where the local net charge density $n$, energy density $e$, pressure $p$
and entropy density $s$ are given by the standard integrals over
the thermal equilibrium distribution function in the local fluid
rest frame and are related by the fundamental thermodynamic relation
\begin{equation}
\label{eq11}
T\,s = p - \mu\, n + e.
\end{equation}
Inserting Eqs.~(\ref{eq5})-(\ref{eq7}) into Eqs.~(\ref{eq1}) and (\ref{eq2}) 
yields the relativistic ideal fluid equations shown in 
Eqs.~(\ref{eq12})-(\ref{eq14}) below. Using Eq.~(\ref{eq11}) together with 
the Gibbs-Duhem relation $dp=s\,dT+n\,d\mu$, it is easy to prove that, 
in the absence of shock discontinuities, these equations also conserve 
entropy, i.e. $\partial_\mu S^\mu=0$.

Note that the validity of the decompositions (\ref{eq5})-(\ref{eq7}) only
requires local momentum isotropy (i.e. that in the local fluid rest frame 
the phase-space distribution reduces to a function of energy $E$ only, 
$f(x,p)=f\bigl(p{\cdot}u(x);T(x),\mu(x)\bigr)$), but not that the 
distribution function has the specific exponential form (\ref{eq4}) 
which maximizes entropy. This may have relevance in situations where 
the time scale for local momentum isotropization is much shorter than 
for thermalization \cite{ALM03,BBW04,RRS04} (i.e. it is much easier 
to change the direction of the particles'\ momenta than their energies), 
with the macroscopic hydrodynamic time scale in between. In this case 
the local microscopic states would not maximize entropy, and the 
relation (\ref{eq11}) would not hold between the quantities $e,\,p,\,n,$ 
and $s$ defined through eqs.~(\ref{eq5})-(\ref{eq10}). Still, they 
would follow ideal fluid dynamical evolution since entropy production 
by microscopic kinetic energy-shifting processes would happen only on 
time scales which are large compared to the macroscopic evolution time 
scales. [Note that in the absence of such a clear separation of time 
scales entropy production can not be neglected during the macroscopic 
evolution, and ideal fluid dynamics must be replaced by dissipative 
fluid dynamics.] 

The ideal fluid equations read (with $\theta\equiv\partial{\cdot}u$ denoting
the local expansion rate)
\begin{eqnarray}
\label{eq12}
&&\dot n = - n\, \theta,
\\
\label{eq13}
&&\dot e = - (e+p)\, \theta,
\\
\label{eq14}
&&{\dot u}^\mu = \frac{\nabla^\mu p}{e+p},
\end{eqnarray}
where we decomposed the partial derivative 
$\partial^\mu=u^\mu D +\nabla^\mu$ into ``longitudinal'' 
and ``transverse'' components $D=u^\nu\partial_\nu$ and 
$\nabla^\mu=\Delta^{\mu\nu}\partial_\nu$, which in the local 
fluid rest frame reduce to the time derivative $\dot f \equiv Df$ and 
spatial gradient $\bm{\nabla}f$. The first two equations describe the 
dilution of the local baryon and energy density due to the local
expansion rate $\theta$, while the third describes the 
acceleration of the fluid by the spatial (in the local frame)
pressure gradients, with the enthalpy $e{+}p$ acting as inertia.
The 5 equations (\ref{eq12})-(\ref{eq14}) for the 6 unknown functions 
$n,\,e,\,p,\,u^\mu$ (remember that $u^\mu u_\mu{\,=\,}1$) must be closed
by supplying an {\em Equation of State (EOS)} $p=p(e,n)$.

\section{Non-ideal fluid decomposition}
\label{sec3}
As the hydrodynamic evolution changes the local energy and baryon density,
microscopic processes attempt to readjust the local phase-space distribution 
to corresponding new local temperatures and chemical potentials. If this
does not happen fast enough, the phase-space distribution will
start to deviate from its local equilibrium form (\ref{eq4}): 
$f(x,p)=f_{\rm eq}\bigl(p{\cdot}u(x);T(x),\mu(x)\bigr) + \delta f(x,p)$.
The optimal values for the (readjusted) local temperature and chemical
potential in the first term are fixed by imposing the ``Landau matching 
conditions'' \cite{LL63}
\begin{eqnarray}
\label{eq15}
&&u_\mu\, \delta T^{\mu\nu} u_\nu = \int \frac{d^3p}{E}\, (u{\cdot}p)^2\,
\delta f(x,p)=0,
\\
\label{eq15a}
&& u_\mu \,\delta N^\mu =  \int \frac{d^3p}{E}\, (u{\cdot}p) \,
\delta f(x,p)=0.
\end{eqnarray}
The remaining deviations $\delta f$ from local equilibrium generate
additional terms in the decompositions of $N^\mu,\,T^{\mu\nu},$ and
$S^\mu$:
\begin{eqnarray}
\label{eq16}
&&N^\mu = N^\mu_{\rm eq} +\delta N^\mu = n\,u^\mu + V^\mu,
\\
\label{eq17}
&&T^{\mu\nu} = T_{\rm eq}^{\mu\nu} + \delta T^{\mu\nu}
           = e\,u^\mu u^\nu -(p+\Pi)\Delta^{\mu\nu}
               + \pi^{\mu\nu}
\qquad
\nonumber\\
&&\qquad\qquad\qquad\qquad\quad
   + W^\mu u^\nu +W^\nu u^\mu,
\\
\label{eq18}
&&S^\mu = S^\mu_{\rm eq} +\delta S^\mu = n\,u^\mu + \Phi^\mu.
\end{eqnarray}
The new terms describe a baryon flow 
$V^\mu{\,=\,}\Delta^{\mu\nu} N_\nu$ 
in the local rest frame, an energy flow 
$W^\mu{\,=\,}\frac{e{+}p}{n} V^\mu{\,+\,}q^\mu$ (where $q^\mu$ is the 
``heat flow vector'') in the local rest frame, the viscous bulk pressure 
$\Pi{\,=\,}{-}\frac{1}{3}\Delta_{\mu\nu}T^{\mu\nu}{\,-\,}p$ (which contributes
to the trace of the energy momentum tensor), the traceless viscous
shear pressure tensor $\pi^{\mu\nu}= T^{\langle\mu\nu\rangle} 
\equiv \left[\frac{1}{2}\left(\Delta^{\mu\sigma}
\Delta^{\nu\tau}{+}\Delta^{\nu\sigma}\Delta^{\mu\tau}\right)-\frac{1}{3}
\Delta^{\mu\nu}\Delta^{\sigma\tau}\right]T_{\tau\sigma}$
(where the expression $\langle\mu\nu\rangle$ is a shorthand
for ``traceless and transverse to $u_\mu$ and $u_\nu$'', as defined
by the projector in square brackets), and an entropy flow vector
$\Phi^\mu$ in the local rest frame.

Inserting the decompositions (\ref{eq15}), (\ref{eq17}) into the conservation
laws (\ref{eq1}), (\ref{eq2}) yields the {\em non-ideal fluid equations}
\begin{eqnarray}
\label{eq19}
&&\dot n = - n\, \theta - \nabla{\cdot}V +V{\cdot}{\dot u},
\\
\label{eq20}
&&\dot e = - (e{+}p{+}\Pi)\, \theta 
           +\pi_{\mu\nu}\nabla^{\left\langle\mu\right.}
                             u^{\left.\nu\right\rangle}
           -\nabla{\cdot}W + 2\, W{\cdot}{\dot u},
\qquad
\\
\label{eq21}
&&(e{+}p{+}\Pi)\,{\dot u}^\mu = \nabla^\mu(p{+}\Pi) 
                             -\Delta^{\mu\nu}\nabla^\sigma\pi_{\nu\sigma}
             +\pi^{\mu\nu}{\dot u}_\nu
\nonumber\\
&&\qquad\qquad\qquad
             -\left[\Delta^{\mu\nu}{\dot W}_\nu + W^\mu\theta 
             +(W{\cdot}\nabla)u^\mu\right].
\end{eqnarray}

The matching conditions (\ref{eq15}) leave the choice of the local 
rest frame velocity $u^\mu$ ambiguous. This ambiguity can be used to
eliminate either $V^\mu$ from Eq.~(\ref{eq16}) (``Eckart frame'', no
baryon flow in the local rest frame \cite{Eckart}), in which case the 
energy flow reduces to the heat flow vector $W^\mu{\,=\,}q^\mu$, or 
$W^\mu$ from Eq.~(\ref{eq17}) (``Landau frame'', no energy flow in the 
local rest frame \cite{LL63}), in which case there is a non-zero baryon 
flow $V^\mu{\,=\,}-\frac{n}{e{+}p}q^\mu$ due to heat conduction in the 
local rest frame. (Intermediate frames are also possible, but yield no 
practical advantage.) For systems with vanishing net baryon number 
(as approximately realized in RHIC collisions) the Eckart frame is 
ill-defined \cite{DG84}, so we will use the Landau frame. In this 
frame, for baryon-free systems with $n{\,=\,}0$ and no heat conduction, 
the non-ideal fluid equations (\ref{eq19})-(\ref{eq21}) simplify to
\begin{eqnarray}
\label{eq22}
&&\!\!\!\!\! \dot e = - (e{+}p{+}\Pi)\, \theta 
           +\pi_{\mu\nu}\nabla^{\left\langle\mu\right.}
                             u^{\left.\nu\right\rangle},
\\
\label{eq23}
&&\!\!\!\!\! (e{+}p{+}\Pi)\,{\dot u}^\mu = \nabla^\mu(p{+}\Pi) 
                             -\Delta^{\mu\nu}\nabla^\sigma\pi_{\nu\sigma}
             +\pi^{\mu\nu}{\dot u}_\nu.
\qquad 
\end{eqnarray}

The non-equilibrium decompositions (\ref{eq16})-(\ref{eq18}) involve
1+3+5=9 additional dynamical quantities, the ``dissipative flows''
$\Pi,\,q^\mu$, and $\pi^{\mu\nu}$ (the counting reflects their 
transversality to $u^\mu$ and the tracelessness of $\pi^{\mu\nu}$). 
This means that we need 9 additional dynamical equations which 
should be compatible with the underlying transport theory for the 
non-equilibrium deviation $\delta f(x,p)$. For the baryon-free case
without heat conduction, the number of needed additional equations
reduces to 6.

\section{Kinetic equations for the dissipative flows}
\label{sec4}
The key property of the kinetic equation governing the evolution of
the phase-space distribution function $f{\,=\,}f_{\rm eq}{+}\delta f$
is that the collision term satisfies the second law of thermodynamics
(\ref{eq3}), i.e. entropy is produced until the system has reached a 
new state of local thermal equilibrium. We don't want to solve the
kinetic theory; instead, we want to write down a phenomenological 
macroscopic theory which is consistent with the constraints arising from
the underlying kinetic theory, in particular the 2$^{\rm nd}$ law. The
macroscopic theory will be constructed from an expansion of the
entropy production rate in terms of the dissipative flows which 
themselves are proportional to the off-equilibrium deviation $\delta f$
of the phase-space distribution \cite{IS79}. Assuming the latter to be small,
$|\delta f|{\,\ll\,}|f_{\rm eq}|$, this expansion will be truncated
at some low order in the dissipative flows $\delta N^\mu,\,\delta
T^{\mu\nu}$. The expansion will involve phenomenological expansion 
coefficients which, in principle, should be matched to the kinetic 
theory \cite{IS79}. In practice, they will often be considered as
phenomenological parameters to be adjusted to experimental data.
In the end, the extracted values must then be checked for consistency 
with the entire approach, by making sure that the dissipative corrections
are indeed sufficiently small to justify truncation of the expansion 
{\em a posteriori}.

The equilibrium identity (\ref{eq11}) can be rewritten as
\begin{equation}
\label{eq24}
S_{\rm eq}^\mu = p(\alpha,\beta)\beta^\mu - \alpha N_{\rm eq}^\mu
+\beta_\nu T_{\rm eq}^{\nu\mu},
\end{equation}
where $\alpha{\,\equiv\,}\frac{\mu}{T},\,\beta{\,\equiv\,}\frac{1}{T},$ and 
$\beta_\nu{\,\equiv\,}\frac{u_\nu}{T}$. The most general off-equilibrium
generalization of this is \cite{IS79}
\begin{eqnarray}
\label{eq25}
S^\mu &\equiv& S_{\rm eq}^\mu + \Phi^\mu
\\\nonumber
&=& p(\alpha,\beta)\beta^\mu - 
\alpha N^\mu +\beta_\nu T^{\nu\mu} + Q^\mu(\delta N^\mu,\delta T^{\mu\nu}),
\end{eqnarray}
where, in addition to the first order contributions implicit in
the second and third terms of the r.h.s., $Q^\mu$ includes terms which
are second and higher order in the dissipative flows $\delta N^\mu$
and $\delta T^{\mu\nu}$. [Note that, by using the identity (\ref{eq11})
between the equilibrium quantities, Eq.~(\ref{eq25}) can be written
in the simpler-looking form $S^\mu{\,=\,}s\,u^\mu{+}\frac{q^\mu}{T}{+}Q^\mu$
but this is not helpful for calculating the entropy production rate.]

The form of the expansion (\ref{eq25}) is constrained by the
2$^{\rm nd}$ law $\partial_\mu S^\mu{\,\geq\,}0$. To evaluate this 
constraint it is useful to rewrite the Gibbs-Duhem relation 
$dp{\,=\,}s\,dT+n\,d\mu$ as
\begin{equation}
\label{eq26}
\partial_\mu\left( p(\alpha,\beta)\beta^\mu\right) = 
N_{\rm eq}^\mu\partial_\mu\alpha - T_{\rm eq}^{\mu\nu}\partial_\mu\beta_\nu.
\end{equation}
With additional help from the conservation laws (\ref{eq1})and (\ref{eq2}), 
the entropy production then becomes
\begin{equation}
\label{eq27}
\partial_\mu S^\mu = - \delta N^\mu\partial_\mu\alpha
+ \delta T^{\mu\nu}\partial_\mu\beta_\nu +\partial_\mu Q^\mu.
\end{equation}
Using Eqs.~(\ref{eq16},\ref{eq17}) to express $\delta N^\mu$ and $\delta 
T^{\mu\nu}$ in terms of the scalar, vector and tensor dissipative flows
$\Pi,\,q^\mu,$ and $\pi^{\mu\nu}$, and introducing corresponding scalar,
vector and tensor thermodynamic forces (in terms of gradients of the
thermodynamic equilibrium variables) which drive these dissipative flows,
$X{\,\equiv\,}{-}\theta{\,=\,}{-}\nabla{\cdot}u$, \ 
$X^\nu{\,\equiv\,}\frac{\nabla^\nu T}{T}-{\dot u}^\nu = {-}\frac{nT}{e{+}p}\,
\nabla^\nu\!\left(\frac{\mu}{T}\right)$, and
$X^{\mu\nu}{\,\equiv\,}\nabla^{\left\langle\mu\right.}
 u^{\left.\nu\right\rangle}$ (note that $X^{\mu\nu}{=}X^{\langle\mu\nu\rangle}$
is traceless and transverse to $u$), the 2$^{\rm nd}$ law constraint 
can be further recast into
\begin{equation}
\label{eq28}
 T \partial_\mu S^\mu = \Pi X - q^\mu X_\mu +\pi^{\mu\nu} X_{\mu\nu}
+ T \partial_\mu Q^\mu \geq 0.
\end{equation}
Note that the first three terms on the r.h.s. are first order while the 
last term is higher order in the dissipative flows.

\subsection{Standard dissipative fluid dynamics (first order theory)}
\label{sec4a}
%
The standard approach (which can be found, for example, in \cite{LL63})
one neglects the higher order contributions and sets $Q^\mu{\,=\,}0$. The 
inequality (\ref{eq28}) can then always be satisfied by postulating 
linear relationships between the dissipative flows and the thermodynamic 
forces,
\begin{subequations}
\label{eq29}
\begin{eqnarray}
\label{eq29a}
  &&\Pi=-\zeta\theta,
\\
\label{eq29b}
  &&q^\nu=-\lambda \frac{nT^2}{e{+}p}\,\nabla^\nu\!\left(\frac{\mu}{T}\right),
\\
\label{eq29c}
  &&\pi^{\mu\nu} = 2\, \eta\, \nabla^{\left\langle\mu\right.}
  u^{\left.\nu\right\rangle},
\end{eqnarray}
\end{subequations}
with positive {\em transport coefficients} $\zeta{\,\geq\,}0$ (bulk
viscosity),  $\lambda{\,\geq\,}0$ (heat conductivity), and 
$\eta{\,\geq\,}0$ (shear viscosity):
\begin{equation}
\label{eq30}
T \partial{\cdot}S = \frac{\Pi^2}{\zeta} - \frac{q^\alpha q_\alpha}{2\lambda T}
+ \frac{\pi^{\alpha\beta}\pi_{\alpha\beta}}{2\eta}\geq 0.
\end{equation}
(The minus sign in front of the second term is necessary because $q^\mu$,
being orthogonal to $u^\mu$, is spacelike, $q^2<0$.) These are the desired 
9 equations for the dissipative flows. 

Unfortunately, using these relations in the hydrodynamic equations 
(\ref{eq19})-(\ref{eq21}) leads to hydrodynamic evolution with acausal 
signal propagation: if in a given fluid cell at a certain time a 
thermodynamic force happens to vanish, the corresponding dissipative
flow also stops instantaneously. This contradicts the fact that the 
flows result from the forces through microscopic scattering which 
involves relaxation on a finite albeit short kinetic time scale.
To avoid this type of acausal behaviour one must keep $Q^\mu$.

\subsection{Second order Israel-Stewart theory}
\label{sec4b}
%
A causal theory of dissipative relativistic fluid dynamics is obtained
by keeping $Q^\mu$ up to terms which are second order in the irreversible 
flows. For simplicity we here consider only the baryon-free case 
$n{\,=\,}q^\mu{\,=\,}0$; for a general treatment see 
\cite{IS79,Muronga:2001zk}. One writes \cite{IS79}
\begin{equation}
\label{eq31}
  Q^\mu = -\left(\beta_0\Pi^2 + \beta_2\pi_{\nu\lambda}\pi^{\nu\lambda}\right)
  \frac{u^\mu}{2T}
\end{equation}
(with phenomenological expansion coefficients $\beta_0,\,\beta_2$) and 
computes (after some algebra using similar techniques as before) the
entropy production rate as
\begin{eqnarray}
\label{eq32}
  T \partial{\cdot}S &=& \Pi \left[ -\theta -\beta_0 \dot\Pi - \Pi T 
  \partial_\mu\left(\frac{\beta_0 u^\mu}{2T}\right)\right]
\\\nonumber
  &+& \pi^{\alpha\beta}\left[\nabla_{\left\langle\alpha\right.}
  u_{\left.\beta\right\rangle} - \beta_2{\dot\pi}_{\alpha\beta}
  -\pi_{\alpha\beta} T \partial_\mu\left(\frac{\beta_2 u^\mu}{2T}\right)
  \right].
\end{eqnarray}
From the expressions in the square brackets we see that the thermodynamic 
forces $-\theta$ and $\nabla_{\left\langle\alpha\right.} 
u_{\left.\beta\right\rangle}$ are now self-consistently modified by
terms involving the time derivatives (in the local rest frame) of the 
irreversible flows $\Pi$, $\pi_{\alpha\beta}$. This leads to dynamical
(``transport'') equations for the latter. We can ensure the 2$^{\rm nd}$ 
law of thermodynamics by again writing the entropy production rate in 
the form (\ref{eq30}) (without the middle term), which amounts to 
postulating 
\begin{eqnarray}
\label{eq33}
\dot \Pi &=& -\frac{1}{\tau_{_\Pi}}\left[ \Pi +\zeta \theta +
\Pi\zeta T \partial_\mu\left(\frac{\tau_{_\Pi} u^\mu}{2\zeta T}\right)\right]
\nonumber\\
 &\approx& -\frac{1}{\tau_{_\Pi}}\bigl[ \Pi +\zeta \theta\bigr],
\\
\label{eq34}
\dot \pi_{\alpha\beta} &=& -\frac{1}{\tau_\pi}\left[ \pi_{\alpha\beta}
- 2 \eta \nabla_{\left\langle\alpha\right.} u_{\left.\beta\right\rangle}
+ \pi_{\alpha\beta} \eta T \partial_\mu\left(\frac{\tau_\pi u^\mu}{2\eta T}
\right)\right]
\nonumber\\
 &\approx& -\frac{1}{\tau_\pi}\left[ \pi_{\alpha\beta}
   - 2 \eta \nabla_{\left\langle\alpha\right.} u_{\left.\beta\right\rangle}
\right].
\end{eqnarray}
Here we replaced the coefficients $\beta_{0,2}$ by the relaxation times 
$\tau_{_\Pi}{\,\equiv\,}\zeta\beta_0$ and $\tau_\pi{\,\equiv\,}2\eta\beta_2$.
In principle both $\zeta,\eta$ and $\tau_{_\Pi},\tau_\pi$ should be calculated
from the underlying kinetic theory. We will use them as phenomenological
parameters, noting that for consistency the microscopic relaxation rates
should be much larger than the local hydrodynamic expansion rate, 
$\tau_{_{\pi,\Pi}}\theta{\,\ll\,}1$.

Let us shortly discuss the approximation in the second equalities
in Eqs.~(\ref{eq33}) and (\ref{eq34}): We are using an expansion scheme for 
the entropy production rate in which the thermodynamic forces and irreversible
flows are assumed to be small perturbations. The approximation in
Eqs.~(\ref{eq33}), (\ref{eq34}) neglects terms which are products of
the irreversible flows with gradients of the thermodynamic equilibrium
quantities which are of the same order as the thermodynamic forces.
These terms are thus effectively of second order in small quantities
and should, for consistency, be neglected relative to the other terms
in the square brackets which are of first order. If one wants to keep
them (as done by Muronga \cite{Muronga:2001zk,MR04}), one should also
keep third-order terms in the entropy flow vector $Q^\mu$ for consistency.
Of course, where the thermodynamic forces and irreversible flows are 
really small, it shouldn't matter whether we keep or drop these terms.
In practice, however, one will use this approach when dissipative effects
are expected to be significant, and the dropped terms may not be extremely
small. In this case we believe that dropping them is more consistent
than keeping them.

There is another reason for dropping these terms: without them, 
Eqs.~(\ref{eq33}), (\ref{eq34}) are relaxation equations which describe
(in the local rest frame) exponential relaxation (on the time scales 
$\tau_{_{\pi,\Pi}}$) of the irreversible flows to the values given
by Eqs.~(\ref{eq29}) in the first order theory. However, if these 
terms are kept, one has instead equations of the form
\begin{eqnarray}
\label{eq35}
\dot \Pi &=& -\frac{1}{\tau_{_\Pi}}\Bigl[ \Pi +\zeta \theta +
\Pi\zeta \gamma_{_\Pi}\Bigr] 
\\
&=& -\frac{1{+}\gamma_{_\Pi}\zeta}{\tau_{_\Pi}}
\left[ \Pi + \frac{\zeta}{1{+}\gamma_{_\Pi}\zeta}\,\theta\right]
= -\frac{1}{\tau'_{_\Pi}}\Bigl[ \Pi +\zeta'\, \theta\Bigr],
\nonumber
\end{eqnarray}
and similarly for the shear pressure tensor. One sees that both the 
kinetic relaxation time and the viscosity are modified by the
factor $\gamma_{_\Pi}{\,=\,}T \partial_\mu\left(\frac{\tau_{_\Pi} 
u^\mu}{2\zeta T}\right)$ which involves the macroscopic expansion 
rate $\partial_\mu u^\mu$. This contradicts the intuitive expectation
that these transport coefficients should
be expressible through integrals of the kinetic collision term
which involve only microscopic physics (cross sections, local densities, 
etc.) 

In the second order Israel-Stewart formalism, one thus solves the
dissipative hydrodynamic equations (\ref{eq19})-(\ref{eq21}) simultaneously
with the kinetic relaxation equations (\ref{eq33}), (\ref{eq34}) for
the irreversible flows. Let us now look at these equations in more 
detail when expressed in a global coordinate system (and not in local
rest frame coordinates as done up to now).

\section{Transverse expansion dynamics in systems with longitudinal 
boost invariance}
\label{sec5}
We are restricting our discussion to systems with longitudinal
boost invariance. With this approximation we can describe the 
transverse expansion in very high energy heavy-ion collisions in
a domain near midrapidity. Boost-invariant systems are conveniently
described in $(\tau,x,y,\eta)$ coordinates where 
$\tau{\,=\,}\sqrt{t^2{-}z^2}$ is longitudinal proper time,
$\eta{\,=\,}\frac{1}{2}\ln[(t{+}z)/(t{-}z)]$ is space-time rapidity, and 
$\bm{r_\perp}{\,=\,}(x,y)$ are the usual Cartesian coordinates in the
plane transverse to the beam direction $z$. Boost-invariant systems 
are then characterized by macroscopic observables which are independent
of $\eta$ and by phase-space distributions which depend only on the difference
$Y{-}\eta$ (where $Y{\,=\,}\frac{1}{2}\ln[(E{+}p_z)/(E{-}p_z)]$ is the
momentum-space rapidity of a particle with longitudinal momentum $p_z$
and energy $E$). We denote 2-dimensional vectors in the transverse plane
by $\bm{r}_\perp,\,\bm{v}_\perp,\,\bm{\nabla}_\perp,$ etc., and generally
use lowercase latin letters to denote vector and tensor components in 
this curvilinear space-time coordinate system. The metric tensor in this 
coordinate system reads
\begin{eqnarray}
\label{eq36}
g^{mn}&=&{\rm diag}\,\bigl(1,-1,-1,-1/\tau^2\bigr),
\nonumber\\
g_{mn}&=&{\rm diag}\,\bigl(1,-1,-1,-\tau^2\bigr).
\end{eqnarray}
%
The flow velocity is parametrized as 
\begin{equation}
\label{eq37}
u^m = \gamma_\perp \bigl(1,v_x,v_y,0\bigr) = \gamma_\perp 
      \bigl(1,\bm{v}_\perp,0\bigr)
\end{equation}
where
\begin{equation}
\label{eq37a}
      \gamma_\perp = \frac{1}{\sqrt{1-v_\perp^2}} 
                   = \frac{1}{\sqrt{1-v_x^2-v_y^2}},
\end{equation}
with vanishing flow component $u^\eta$ in $\eta$ direction
and transverse flow velocity $\bm{v}_\perp(\tau,\bm{r}_\perp)$.
For vectors and tensors, the usual Cartesian derivatives 
$\partial_\mu$ must be replaced by covariant derivatives, denoted 
by semicolons:
\begin{subequations}
\label{eq38}
\begin{eqnarray}
\label{eq38a}
&&\partial_\mu j^\nu \to j^n_{\ ;m} = \partial_m j^n
+ \Gamma^n_{mk} j^k, 
\\
\label{eq38b}
&&\partial_\mu T^{\nu\lambda} \to T^{nl}_{\ \ ;m} =
\partial_m T^{nl} + \Gamma^n_{mk} T^{kl} + T^{nk} \Gamma^l_{km},
\qquad\ \ 
\end{eqnarray}
\end{subequations}
where $\Gamma^i_{jk}{\,=\,}\frac{1}{2}g^{im}\bigl(\partial_j 
g_{km}{+}\partial_k g_{mj}{-}\partial_m g_{jk}\bigr)$ are
the Chri\-stof\-fel symbols. The only nonvanishing components
of $\Gamma^i_{jk}$ are
\begin{equation}
\label{eq39}
\Gamma^\eta_{\eta\tau} = \Gamma^\eta_{\tau\eta} = \frac{1}{\tau},
\quad  \Gamma^\tau_{\eta\eta} = \tau.
\end{equation}
The time derivative in the local comoving frame and the local
expansion rate are thus computed as
\begin{eqnarray}
\label{eq40}
D &=& u\cdot\partial = \gamma_\perp \bigl(\partial_\tau + 
      \bm{v}_\perp{\cdot}\bm{\nabla}_\perp\bigr),
\\
\label{eq41}
\theta &=& \partial\cdot u = \frac{1}{\tau}\, 
           \partial_\tau\left(\tau\gamma_\perp\right) 
         + \bm{\nabla}_\perp{\cdot}\left(\gamma_\perp \bm{v}_\perp\right).
\end{eqnarray}

If the expanding system has additionally azimuthal symmetry around the
beam direction (for example, central collisions between spherically 
symmetric nuclei), it is advantageous to replace the Cartesian transverse
coordinates $(x,y)$ by polar coordinates $(r,\phi)$ since macroscopic 
quantities are then $\phi$-independent:
\begin{subequations}
\label{eq42}
\begin{eqnarray}
\label{eq42a}
&&x^m = (\tau,r,\phi,\eta),
\\
\label{eq42b}
&&g^{mn}={\rm diag}\,\bigl(1,-1,-1/r^2,-1/\tau^2\bigr),
\\
\label{eq42c}
&&g_{mn}={\rm diag}\,\bigl(1,-1,-r^2,-\tau^2\bigr).
\end{eqnarray}
\end{subequations}
This leads to the following additional (to Eq.~(\ref{eq39}))
non-vanishing Christoffel symbols
\begin{equation}
\label{eq43}
\Gamma^\phi_{\phi r} = \Gamma^\phi_{r\phi} = +\frac{1}{r},
\quad  \Gamma^r_{\phi\phi} = - r.
\end{equation}
The flow velocity now simplifies to
\begin{equation}
\label{eq44}
u^m = \gamma_r \bigl(1,v_r,0,0\bigr) \quad \mbox{with} \quad
      \gamma_r = \frac{1}{\sqrt{1-v_r^2}} 
\end{equation}
with radial transverse flow velocity $\bm{v}_\perp{\,=\,}v_r(\tau,r)\,\bm{e}_r$
and vanishing flow components $u^\phi$ and $u^\eta$. Correspondingly
the time derivative in the local comoving frame and the local
expansion rate reduce to
\begin{eqnarray}
\label{eq45}
&&D = u\cdot\partial = \gamma_r \bigl(\partial_\tau + v_r \partial_r\bigr),
\\
\label{eq46}
&&\theta = \partial\cdot u 
        = \frac{1}{\tau}\, \partial_\tau\left(\tau\gamma_r\right) 
        + \frac{1}{r}\, \partial_r\left(r v_r \gamma_r\right).
\qquad
\end{eqnarray}
We will now treat the azimuthally symmetric and non-symmetric cases
separately.

\subsection{(1+1)-dimensional viscous hydrodynamics: the azimuthally
symmetric case}
\label{sec5a}
%
Due to azimuthal symmetry and longitudinal boost invariance, the
$n{\,=\,}\phi$ and $n{\,=\,}\eta$ components of the equations of motion
${T^{mn}}_{;m}{\,=\,}0$ are redundant. Using the results of 
Appendix~\ref{appa}, the $n{\,=\,}\tau$ and $n{\,=\,}r$ components can 
be written as
\begin{eqnarray}
\label{eq47}
&&\frac{1}{\tau}\partial_\tau\Bigl(\tau T^{\tau\tau}\Bigr) +
  \frac{1}{r}\partial_r\Bigl(r T^{\tau r}\Bigr) 
\nonumber\\
&&\qquad\qquad\qquad\qquad
  = -\,\frac{p+\Pi+\tau^2 \pi^{\eta\eta}}{\tau},
\\
\label{eq48}
&&\frac{1}{\tau}\partial_\tau\Bigl(\tau T^{\tau r}\Bigr) +
  \frac{1}{r}\partial_r\Bigl(r (T^{\tau r} v_r + {\cal P}_r)\Bigr) 
\nonumber\\
&&\qquad\qquad\qquad\qquad
= +\,\frac{p+\Pi+r^2\pi^{\phi\phi}}{r}.
\qquad
\end{eqnarray}
%
With the shorthand notations $\tilde T^{mn}{\,=\,}r \tau T^{mn}$,
$\tilde{\cal P}_r{\,=\,}r \tau {\cal P}_r$, and 
$\tilde v_r{\,=\,}\frac{\tilde T^{\tau r}}{\tilde T^{\tau\tau}}{\,=\,}\frac
{T^{\tau r}}{T^{\tau\tau}}$ these are
brought into ``standard (Cartesian) form''
\begin{eqnarray}
\label{eq49}
&&\partial_\tau \tilde T^{\tau\tau} 
+ \partial_r(\tilde v_r \tilde T^{\tau\tau}) 
= -r \Bigl(p+\Pi+\tau^2 \pi^{\eta\eta}\Bigr),
\qquad\qquad
\\
&&\partial_\tau \tilde T^{\tau r} + \partial_r\Bigl(v_r \tilde T^{\tau r} 
  + \tilde {\cal P}_r\Bigr) = \tau\Bigl(p+\Pi+r^2\pi^{\phi\phi}\Bigr).
\nonumber
\end{eqnarray}
The corresponding transport equations for the dissipative fluxes read
(using the explicit expressions (\ref{A11}) for the shear tensor from
Appendix~\ref{appa})
\begin{eqnarray}
\label{eq51}
&&\Bigl(\partial_\tau + v_r\partial_r\Bigr)\pi^{\eta\eta} = 
  -\frac{1}{\gamma_r\tau_\pi}\left[\pi^{\eta\eta}
  -\frac{2\eta}{\tau^2}\left(\frac{\theta}{3}-\frac{\gamma_r}{\tau}
   \right)\right],
\nonumber\\
\label{eq52}
&&\Bigl(\partial_\tau + v_r\partial_r\Bigr)\pi^{\phi\phi} = 
  -\frac{1}{\gamma_r\tau_\pi}\left[\pi^{\phi\phi}
  -\frac{2\eta}{r^2}\left(\frac{\theta}{3}-\frac{\gamma_r v_r}{r}
   \right)\right],
\nonumber\\
\label{eq53}
&&\Bigl(\partial_\tau + v_r\partial_r\Bigr)\Pi = 
  -\frac{1}{\gamma_r\tau_{_\Pi}}\left[\Pi + \zeta\theta\right].
\end{eqnarray}
Similar equations were derived in \cite{MR04} (with extra terms, 
however, resulting from the higher order corrections in Eqs.~(\ref{eq33}) 
and (\ref{eq34}) which we argued should be neglected). These equations 
can be solved with the code LCPFCT \cite{LCPFCT}, using subroutine 
{\it LCPFCT} for equations (\ref{eq49}) and subroutine 
{\it CNVFCT} for equations (\ref{eq51}). First attempts 
at a numerical solution have been reported in Refs.~\cite{MR04,CH05}, 
but a number of open questions remain at this point, and we leave a 
detailed discussion of the numerical aspects to a subsequent publication. 

The hydrodynamic equations require the equation of state $p(e)$
for closure, i.e. after each transport step in time we must extract
at each spatial grid point the boost velocity $v_r$ between the global
and local rest frames and the local energy density $e$ from the dynamical
variables $T^{\tau\tau}$ and $T^{\tau r}$. Equations (\ref{A7})
give the energy density as
\begin{equation}
\label{eq54}
e = T^{\tau\tau} - v_r T^{\tau r},
\end{equation}
where the radial velocity $v_r$ must be extracted from the implicit
equation 
\begin{equation}
\label{eq55}
v_r = \frac{T^{\tau r}}{T^{\tau\tau}+p(e{=}T^{\tau\tau}{-}v_r T^{\tau r})
      + \Pi - r^2 \pi^{\phi\phi} - \tau^2 \pi^{\eta\eta}}
\end{equation}
by a one-dimensional zero search. This is still the same degree of 
numerical complexity as in the ideal fluid case \cite{Rischke};
we will see in the next subsection, however, that this part of the 
problem becomes numerically more involved for dissipative 
hydrodynamics {\it without} azimuthal symmetry.

\begin{widetext}
\subsection{(2+1)-dimensional viscous hydrodynamics with
longitudinal boost invariance}
\label{sec5b}
In the absence of azimuthal symmetry, only the $n{\,=\,}\eta$ component 
of the equations of motion ${T^{mn}}_{;m}{\,=\,}0$ is redundant
(due to boost invariance). Using the relations (\ref{A31})
from Appendix~\ref{appb}, the $n{\,=\,}\tau,x,y$ components can be written as
\begin{subequations}
\label{eq56}
\begin{eqnarray}
\label{eq56a}
&&\frac{1}{\tau}\partial_\tau\Bigl(\tau T^{\tau\tau}\Bigr) 
 +\partial_x T^{\tau x}+\partial_y T^{\tau y} = 
  -\,\frac{p+\Pi+\tau^2 \pi^{\eta\eta}}{\tau},
\\
\label{eq56b}
&&\frac{1}{\tau}\partial_\tau\Bigl(\tau T^{\tau x}\Bigr) 
 +\partial_x \Bigl((T^{\tau x}{-}\pi^{\tau x})v_x\Bigr)
 +\partial_y \Bigl((T^{\tau x}{-}\pi^{\tau x})v_y\Bigr) 
 = -\partial_x(p{+}\Pi{+}\pi^{xx}) - \partial_y\pi^{xy},
\\
\label{eq56c}
&&\frac{1}{\tau}\partial_\tau\Bigl(\tau T^{\tau y}\Bigr) 
 +\partial_x \Bigl((T^{\tau y}{-}\pi^{\tau y})v_x\Bigr)
 +\partial_y \Bigl((T^{\tau y}{-}\pi^{\tau y})v_y\Bigr) 
 = - \partial_x\pi^{xy}-\partial_y(p{+}\Pi{+}\pi^{yy}).
\end{eqnarray}
\end{subequations}
The further manipulation of these equations depends on our choice
of independent shear pressure components as dynamical variables. 
In the following two subsections we explore two different choices, 
each with its own advantages and disadvantages.

\subsubsection{Choosing $\pi^{\tau\tau}$, $\Delta{\,=\,}\pi^{xx}{-}\pi^{yy}$,
and $\pi^{\eta\eta}$ as independent dynamical variables}
\label{sec5b1}
In the first approach we select $\pi^{\tau\tau}$, $\pi^{\eta\eta}$,
and the difference $\Delta{\,=\,}\pi^{xx}{-}\pi^{yy}$ as independent
dynamical components of the shear pressure tensor. The last choice has 
the advantage that it vanishes in the azimuthally symmetric case, 
thereby automatically reducing the number of independent dynamical
variables. The choice of $\pi^{\tau\tau}$ instead of the orthogonal
combination $\Sigma{\,=\,}\pi^{xx}{+}\pi^{yy}$ is a matter of taste
and not essential. They are related by Eq.~(\ref{A29}),
$\pi^{\tau\tau}{\,=\,}\Sigma + \tau^2 \pi^{\eta\eta}$.

Introducing the shorthand notations $\tilde T^{mn}{\,=\,}\tau T^{mn}$
(without the factor $r$ this time), $\tilde\pi^{mn}{\,=\,}\tau\pi^{mn}$,
$\tilde p{\,=\,}\tau p$, $\tilde \Pi{\,=\,}\tau \Pi$, 
$\tilde v_i{\,=\,}\frac{\tilde T^{\tau i}}{\tilde 
T^{\tau\tau}}{\,=\,}\frac{T^{\tau i}}{T^{\tau\tau}}$ for $i{\,=\,}x,y$, 
and following the procedure of Appendix~\ref{appb}, equations 
(\ref{eq56}) can be recast into ``standard form''
\begin{subequations}
\label{eq59}
\begin{eqnarray}
\label{eq59a}
&&\partial_\tau\tilde T^{\tau\tau} 
 +\partial_x(\tilde T^{\tau x}\tilde v_x)
 +\partial_y(\tilde T^{\tau y}\tilde v_y) = 
  -(p+\Pi+\tau^2 \pi^{\eta\eta}),
\\
\label{eq59b}
&&\partial_\tau\tilde T^{\tau x}
 +\partial_x \Bigl((\tilde T^{\tau x}{-}\tilde\pi^{\tau x})v_x\Bigr)
 +\partial_y \Bigl((\tilde T^{\tau x}{-}\tilde\pi^{\tau x})v_y\Bigr)
 = -\partial_x(\tilde p{+}\tilde \Pi{+}\tilde\pi^{xx})
   -\partial_y\tilde\pi^{xy},
\\
\label{eq59c}
&&\partial_\tau\tilde T^{\tau y}
 +\partial_x \Bigl((\tilde T^{\tau y}{-}\tilde\pi^{\tau y})v_x\Bigr)
 +\partial_y \Bigl((\tilde T^{\tau y}{-}\tilde\pi^{\tau y})v_y\Bigr) 
 = -\partial_x\tilde\pi^{xy}
   -\partial_y(\tilde p{+}\tilde\Pi{+}\tilde\pi^{yy}),
\end{eqnarray}
\end{subequations}
\end{widetext}
where we have resisted to insert the lengthy explicit expressions 
(\ref{A32}) for $\pi^{\tau x}$, $\pi^{\tau y}$, $\pi^{xy}$, 
$\pi^{xx}$, and $\pi^{yy}$. Note that the latter involve the velocities 
$v_x$ and $v_y$, so in order to evaluate the sources at time step $n$
which drive the propagation to time step $n{+}1$ we must explicitly
solve for the velocities $v_{x,y}(\bm{r}_\perp)$ at time step $n$.
We will return to this issue momentarily.

The transport equations to be solved together with 
Eqs.~(\ref{eq59}) are (see Appendix\ref{appb})
\begin{subequations}
\label{eq62}
\begin{eqnarray}
\label{eq62a}
&&\Bigl(\partial_\tau + v_x\partial_x + v_y\partial_y\Bigr)\pi^{\eta\eta}
  = -\frac{1}{\gamma_\perp\tau_\pi}\Bigl(\pi^{\eta\eta}
    - 2\eta \sigma^{\eta\eta}\Bigr),
\quad\nonumber\\
\\
\label{eq62b}
&&\Bigl(\partial_\tau + v_x\partial_x + v_y\partial_y\Bigr)\pi^{\tau\tau}
  = -\frac{1}{\gamma_\perp\tau_\pi}\Bigl(\pi^{\tau\tau}
    - 2\eta \sigma^{\tau\tau}\Bigr),
\nonumber\\ \\
\label{eq62c}
&&\Bigl(\partial_\tau + v_x\partial_x + v_y\partial_y\Bigr)\Delta
  = -\frac{1}{\gamma_\perp\tau_\pi}\Bigl(\Delta
    - 2\eta \sigma^{\Delta}\Bigr),
\qquad\ \\
\label{eq62d}
&&\Bigl(\partial_\tau + v_r\partial_r\Bigr)\Pi = 
  -\frac{1}{\gamma_\perp\tau_{_\Pi}}\Bigl(\Pi + \zeta\theta\Bigr),
\end{eqnarray}
\end{subequations}
where $\sigma^{\eta\eta},\,\sigma^{\tau\tau}$, and $\sigma^\Delta$ are 
given by Eqs.~(\ref{A37}).

In order to compute the pressure $p$ from the equation of state $p(e)$,
we calculate from the dynamical variables $T^{\tau\tau}$, $T^{\tau x}$,
and $T^{\tau y}$ the energy density by combining 
Eqs.~(\ref{A31a})-(\ref{A31c}):
\begin{equation}
\label{eq66}
e = T^{\tau\tau} - v_x T^{\tau x} - v_y T^{\tau y}.
\end{equation}
(Note that all viscous pressures cancel in this relation.) This requires 
the velocities $v_x$ and $v_y$ which are given (implicitly!) by
\begin{subequations}
\label{eq67}
\begin{eqnarray}
\label{eq67a}
&&v_x =\frac{T^{\tau x} -\pi^{\tau x}}{T^{\tau\tau}+p(e)+\Pi-\pi^{\tau\tau}},
\\
\label{eq67b}
&&v_y =\frac{T^{\tau y} -\pi^{\tau y}}{T^{\tau\tau}+p(e)+\Pi-\pi^{\tau\tau}}.
\end{eqnarray}
\end{subequations}
Since the vectors $(T^{\tau x},T^{\tau y})$ and $(\pi^{\tau x},\pi^{\tau y})$
are not parallel to each other, the direction of the flow velocity
$\bm{v}_\perp{\,=\,}(v_x,v_y)$ is no longer given by the direction
of $(T^{\tau x},T^{\tau y})$ as is the case in ideal fluid dynamics
\cite{Rischke}, and Eqs.~(\ref{eq67}) can no longer be reduced to
a simple one-dimensional zero search. Instead one must simultaneously 
iterate two equations, one for the magnitude of the transverse velocity,
\begin{equation}
\label{eq68}
v_\perp^2 = \frac{T^{\tau\tau}-\pi^{\tau\tau}-e}
                 {T^{\tau\tau}-\pi^{\tau\tau}+p(e)+\Pi}
\end{equation}
(which is easily verified from Eq.~(\ref{A31a})), and one for the 
azimuthal angle of the velocity vector, 
$\phi_v=\tan^{-1}\left(\frac{v_y}{v_x}\right)$.
The latter is obtained by writing 
$v_x=v_\perp\cos\phi_v\equiv v_\perp z$,
$v_y{\,=\,}v_\perp\sin\phi_v{\,=\,}v_\perp\sqrt{1{-}z^2}$, dividing 
the two equations (\ref{eq67}) by each other and inserting 
Eqs.~(\ref{A32a}), (\ref{A32b}):
\begin{widetext}
\begin{eqnarray}
\label{eq69}
z^2\biggl[2v_\perp\sqrt{1{-}z^2}\,T^{\tau y}&-&\left.\pi^{\tau\tau}
  +\frac{v_\perp^2}{2}\left((\pi^{\tau\tau}{-}\tau^2\pi^{\eta\eta})(2z^2{-}1)
  +\frac{\Delta}{2}\right)\right]
\nonumber\\
  -(1{-}z^2)\biggl[2v_\perp z\, T^{\tau x}&-&\left.\pi^{\tau\tau}
  -\frac{v_\perp^2}{2}\left((\pi^{\tau\tau}{-}\tau^2\pi^{\eta\eta})(2z^2{-}1)
  +\frac{\Delta}{2}\right)\right] =0.
\end{eqnarray}
\end{widetext}
Since Eq.~(\ref{eq68}) requires knowledge of the direction of $\bm{v}_\perp$ 
on the right hand side because of Eq.~(\ref{eq66}), and Eq.~(\ref{eq69})
requires knowledge of $v_\perp$, these two equations cannot be decoupled,
and the iteration problem is genuinely 2-dimensional. This is of serious
concern since this problem must be solved at every spatial grid point 
after each time step which makes it numerically very expensive.

\subsubsection{Choosing $\pi^{\tau\tau}$, $\pi^{\tau x}$, $\pi^{\tau y}$, 
and $\pi^{\eta\eta}$ as independent dynamical variables}
\label{sec5b2}
This problem can be avoided if, instead of $\Delta$, $\pi^{\tau x}$ and 
$\pi^{\tau y}$ are kept as dynamical variables which are directly evolved 
in time via their own kinetic transport equations. Defining the 
2-dimensional vector $\bm{M}{\,=\,}(M_x,M_y)\equiv
(T^{\tau x}{-}\pi^{\tau x},T^{\tau y}{-}\pi^{\tau y})$
(this is just the transverse momentum density vector {\em without}
the shear pressure contributions), we see from Eqs.~(\ref{eq67}) that
$\bm{v}_\perp$ and $\bm{M}$ are parallel, 
$\bm{v}_\perp{\cdot}\bm{M}{\,=\,}v_\perp M$ with 
$M{\,=\,}\sqrt{M_x^2{+}M_y^2}$. 
Introducing furthermore $M_0{\,\equiv\,}T^{\tau\tau}{-}\pi^{\tau\tau}$, 
Eq.~(\ref{eq66}) can with the help of Eq.~(\ref{A30c}) be 
rewritten as
\begin{equation}
\label{eq70}
e = M_0 - \bm{v}_\perp{\cdot}\bm{M} = M_0 - v_\perp M
\end{equation}
which requires knowledge of only the magnitude of $v_\perp$. The latter
can then be obtained by a normal 1-dimensional zero search from
Eq.~(\ref{eq68}) which can also be written as
\begin{equation}
\label{eq71}
v_\perp = \frac{M}{M_0 + p(e{=}M_0{-}v_\perp M) + \Pi},
\end{equation}
and the velocity components are reconstructed from
\begin{equation}
\label{eq72}
v_x = v_\perp \frac{M_x}{M}, \qquad v_x = v_\perp \frac{M_y}{M}.
\end{equation}

Note that this procedure requires direct knowledge of 
$\pi^{\tau\tau}$, $\pi^{\tau x}$, and $\pi^{\tau y}$ at all spatial
grid points at each time step, so $\pi^{\tau x}$ and $\pi^{\tau y}$
cannot be computed from the constraints (\ref{A32a}), (\ref{A32b}) (for
which the velocities $v_x$ and $v_y$ would need to be already known).
On the other hand, $\pi^{\tau\tau}$, $\pi^{\tau x}$, and $\pi^{\tau y}$
are not independent, but related by Eq.~(\ref{A30c}).
Since the suggested procedure requires propagating all three of these
shear pressure components independently via kinetic transport 
equations (which, of course, should accurately preserve the constraint 
(\ref{A30c}) if correctly implemented numerically), we have to solve one 
more kinetic transport equation (involving one physically redundant 
component) than in the procedure of Section~\ref{sec5b1}: Instead of 
the three independent kinetic transport equations 
(\ref{eq62a})-(\ref{eq62c}) we have to solve four equations for 
$\pi^{\tau\tau}$, $\pi^{\tau x}$, $\pi^{\tau y}$, and $\pi^{\eta\eta}$.

The set of equations to be solved simultaneously in this approach is
given by Eqs.~(\ref{eq59}), (\ref{eq62}a,b,d),
plus the following two equations:
\begin{eqnarray}
\label{eq73}
&&\Bigl(\partial_\tau + v_x\partial_x + v_y\partial_y)\pi^{\tau x}
  = -\frac{1}{\gamma_\perp\tau_\pi}\Bigl(\pi^{\tau x}
    - 2\eta \sigma^{\tau x}\Bigr),
\quad
\nonumber\\
\\
\label{eq74}
&&\Bigl(\partial_\tau + v_x\partial_x + v_y\partial_y)\pi^{\tau y}
  = -\frac{1}{\gamma_\perp\tau_\pi}\Bigl(\pi^{\tau y}
    - 2\eta \sigma^{\tau y}\Bigr).
\nonumber
\end{eqnarray}
The shear tensor components required here are given in Appendix \ref{appb}, 
Eqs.~(\ref{A43}).

Is this approach more economical than the two-dim\-en\-sional zero search
from the previous subsection? We believe so. If a one-dimensional zero 
search requires $N$ iterations, each with $K$ algebraic manipulations, a 
two-dimensional zero-search would require ${\cal O}(N^2{\cdot}K{\cdot}K')$
algebraic manipulations at each spatial grid point and time step.
Solving instead an additional kinetic transport equation for, say, 
$\pi^{\tau x}$ requires ${\cal O}(K_s{\cdot}K_t)$ algebraic manipulations 
at each time and grid point, where $K_s$ is the number of algebraic steps 
required to evaluate the source $\sigma^{\tau x}$ and $K_t$ is the number 
of algebraic steps involved in the time evolution algorithm. Taking the 
number $K'$ of manipulations required to evaluate Eq.~(\ref{eq69}) to be 
comparable to $K_s$, solving the extra kinetic transport equation should 
be numerically less expensive since we expect $K_t$ to be significantly 
smaller than $N^2{\cdot}K$. The actual numerical implementation will tell 
whether this expectation is borne out. In any case, dissipative hydrodynamics
is considerably more expensive than ideal fluid dynamics, and
efficient coding will be required.  

\section{Summary}
\label{sec6}
%
In this paper we derived explicit equations of motion, in a form that
makes them directly amenable to publicly available transport algorithms
\cite{LCPFCT}, for a causal theory of dissipative hydrodynamic evolution
for relativistic viscous fluids such as those created in relativistic 
heavy-ion collisions. In doing so we followed the pioneering work by 
Israel and Stewart \cite{IS79} which was recently brought to wider
attention and worked out in greater detail by Muronga \cite{Muronga:2001zk}.
Our treatment is still not completely general in that it continues to
assume boost-invariant expansion along the beam direction (thereby 
reducing the spatial dimensionality of the problem by one), but
it goes beyond the existing literature \cite{Muronga:2001zk,Teaney:2004qa}
by allowing for arbitrary transverse expansion, without the additional 
restriction of azimuthal symmetry around the beam direction. It thus 
provides the formal basis for a numerical calculation of elliptic flow 
in relativistic heavy-ion collisions including dissipative effects. 
Such calculations will be needed for the phenomenological determination 
of the viscosity of the quark-gluon plasma from heavy-ion collision data.

We have also added to the discussion in \cite{MR04} of azimuthally 
symmetric (1+1)-dimensional viscous hydrodynamics by reanalyzing the
Israel-Stewart approach \cite{IS79} as presented in Ref.~\cite{Muronga:2001zk}
and pointing out a systematic issue with the expansion of the entropy
current in terms of higher order terms in the dissipative fluxes.
The practical importance of the improved truncation scheme suggested
here will only be fully assessed once the numerical implementation of
our equations \cite{CH05} has been thoroughly tested and becomes 
available for systematic investigations. 

The simplification of the problem resulting from azimuthal symmetry,
by exploiting polar coordinates in the transverse plane, is significant.
The effects of viscosity can be subsumed into an effective radial
pressure, leaving the diagonal structure of the energy-momentum
tensor in the local rest frame intact. As a consequence, the numerically
critical problem of extracting at each time step from the dynamical 
components of $T^{\mu\nu}$ the local flow velocity and energy density
in order to compute the pressure from the equation of state remains
one-dimensional, i.e. of the same complexity as for ideal fluids.
The additional complexity resulting from dissipation thus resides
entirely in the need for solving, together with the two hydrodynamic
evolution equations, three additional kinetic transport equations
for the bulk viscous pressure and for two components of the shear viscous
pressure.

For the general situation without azimuthal symmetry, the use of polar 
coordinates (with their coordinate singularity at $r_\perp{\,=\,}0$) only
complicates matters. It provides no help towards solving the now in general
two-dimensional selfconsistency problem associated with the extraction
of the flow velocity and local energy density from the dynamical variables.
We therefore use Cartesian coordinates in the transverse plane, as has
been the tradition in (2+1)-dimensional ideal fluid dynamics. Unfortunately,
this choice eliminates the possibility of coding the equations in
such a way that the code would automatically take full advantage of
all the simplifications resulting from azimuthal symmetry when handed
an azimuthally symmetric problem. Azimuthally symmetric (1+1)-dimensional   
expansion and asymmetric (2+1)-dimensional expansion require differently
optimized algorithms.

Using Cartesian transverse coordinates, we found a nice way of avoiding 
the above-mentioned two-dim\-en\-sional nature of the numerically critical 
iteration problem for the local energy density, by increasing the set 
of kinetic transport equations not by 1, but by 2 relative to the 
azimuthally symmetric case. By keeping one of the redundant components 
of the shear pressure tensor as a dynamical variable, we can again bring 
the iteration problem for the local energy density into scalar form. We 
believe that the expense for solving an additional transport equation, 
although not negligible, is less than that required for coping with a 
two-dimensional iteration problem at each time step at all spatial 
grid points.

Compared to (2+1)-dimensional ideal fluid dynamics, dissipative dynamics
generates more complicated source terms for the three independent 
hydrodynamic evolution equations and requires the additional simultaneous
solution of {\em five} kinetic transport equations, one for the bulk viscous
pressure and four for shear viscous pressure components (one of them
being physically, but not algorithmically redundant). Altogether the
resulting increase in numerical complexity (compared to the ideal fluid
case) is probably less than an order of magnitude. Given the increase in 
computer speed and power during the past decade, this should be manageable.
     
\acknowledgments
The work of U.H. was supported by the U.S. Department of Energy under 
contract DE-FG02-01ER41190.

\appendix*
\section{Shear tensor and viscous pressure tensor components}
\subsection{Azimuthally symmetric systems}
\label{appa}
For azimuthally symmetric systems we use polar coordinates in the 
transverse plane: $x^m{\,=\,}(\tau,r,\phi,\eta)$. In this coordinate 
system, the global frame and the local fluid rest frame are connected by
a radial boost with velocity $\bm{v}=v_r(r,\tau)\,\bm{e}_r$. Introducing
the fluid rapidi\-ty $y_r{\,=\,}\tanh^{-1}v_r$ such that 
$\gamma_r{\,=\,}\cosh y_r,\,\gamma_r v_r=\sinh v_r$, the 
corresponding Lorentz transformation matrix is given by
\begin{equation}
\label{A1}
{\Lambda^m}_n(v_r) = 
\left(\begin{array}{cccc}
             \cosh y_r & \sinh y_r & 0 & 0\\ 
             \sinh y_r & \cosh y_r & 0 & 0\\ 
                0      &    0      & 1 & 0\\
                0      &    0      & 0 & 1
      \end{array}
\right).
\end{equation}
The projector transverse to the flow vector $u^m$ takes the form
\begin{equation}
\label{A1a}
{\Delta^m}_n =  
\left(\begin{array}{cccc}
            -\gamma_r^2 v_r^2 & \gamma_r^2 v_r & 0 & 0\\ 
            -\gamma_r^2 v_r   & \gamma_r^2  & 0 & 0\\ 
                0      &    0      & 1 & 0\\
                0      &    0      & 0 & 1
      \end{array}
\right)
\end{equation}
(note that this differs from ${\Delta_m}^n$!). 

Due to azimuthal symmetry and longitudinal boost-invariance,
all mixed components involving indices $\phi$ or $\eta$ of the 
stress tensor 
\begin{equation}
\label{A2} 
\sigma^{mn} \equiv 
\nabla^{\left\langle m\right.} u^{\left.n\right\rangle}
\end{equation}
and the shear pressure tensor $\pi^{mn}$ vanish: 
\begin{eqnarray}
\label{A3} 
&&\sigma^{\phi\tau}=\sigma^{\phi r}=\sigma^{\phi\eta}=\sigma^{\eta\tau}=
\sigma^{\eta r}=0,
\nonumber\\
&&\pi^{\phi\tau}=\pi^{\phi r}=\pi^{\phi\eta}=\pi^{\eta\tau}=
\pi^{\eta r}=0.
\end{eqnarray}
This leaves only two independent components for the shear pressure 
tensor which is constrained by the conditions of tracelessness
\begin{equation}
\label{A4} 
\pi^{\tau\tau}=\pi^{rr}+ r^2 \pi^{\phi\phi} + \tau^2 \pi^{\eta\eta}
\end{equation}
and of orthogonality to $u^m{\,=\,}\gamma_r(1,v_r,0,0)$:
\begin{equation}
\label{A5} 
\pi^{\tau\tau} = v_r \pi^{\tau r},
\qquad
\pi^{r\tau} = v_r \pi^{rr}.
\end{equation}
Equations (\ref{A5}) result from the $n{\,=\,}\tau$ and $r$ components of 
$u_m\pi^{mn}{\,=\,}0$; the other two components yield redundant
equations. Equations (\ref{A4}), (\ref{A5}) can be combined to
yield 
\begin{equation}
\label{A6} 
\pi^{rr} = -\gamma_r^2 (r^2 \pi^{\phi\phi} + \tau^2 \pi^{\eta\eta}).
\end{equation}
Identical relations hold between the corresponding components 
of the shear tensor $\sigma^{mn}$. Equations (\ref{A4}) and 
(\ref{A6}) can be used to eliminate $\pi^{\tau\tau}$, $\pi^{r\tau}$ 
and $\pi^{rr}$ in favor of $\pi^{\phi\phi}$ and $\pi^{\eta\eta}$ 
which we keep as independent dynamical variables.
Equation (\ref{A1}) shows that these components are not affected
by the radial boost, i.e. the $\phi\phi$ and $\eta\eta$ components
of the shear pressure tensor are identical in the global frame and in 
the local fluid rest frame. The same is not true for the other non-zero
components of the shear pressure tensor. 

This observation leads to a particularly simple structure of the 
energy-momentum tensor in the global frame, including viscous 
pressure terms. Using Eqs.~(\ref{A4})-(\ref{A6}) in the expressions
for $T^{\tau\tau}$ and $T^{\tau r}$ (which are the two independent 
components being evolved by the hydrodynamic evolution equations) we 
find 
\begin{subequations}
\label{A7}
\begin{eqnarray}
\label{A7a}
T^{\tau\tau} &=& (e{+}p{+}\Pi)\gamma_r^2 - (p{+}\Pi) + \pi^{\tau\tau}
\nonumber\\
             &=& (e+{\cal{P}}_r)\gamma_r^2 - {\cal{P}}_r,
\\
\label{A7b}
T^{\tau r} &=& (e{+}p{+}\Pi)\gamma_r^2 v_r + \pi^{\tau r}
              = (e+{\cal{P}}_r)\gamma_r^2 v_r,
\qquad
\end{eqnarray}
\end{subequations}
where 
\begin{equation}
\label{A9}
{\cal{P}}_r = p + \Pi - r^2 \pi^{\phi\phi} - \tau^2 \pi^{\eta\eta}.
\end{equation}
Equations (\ref{A7}) are the same expressions as for the ideal fluid, 
except for the replacement of $p$ by the {\em effective radial pressure} 
${\cal{P}}_r$ which has no explicit dependence on the flow velocity $v_r$.

In the local rest frame, the energy-momentum tensor is diagonal:
\begin{equation}
\label{A10}
\hat T^{mn} = 
\left(\begin{array}{cccc}
          e & 0 & 0 & 0\\ 
          0 & {\cal P}_r & 0 & 0\\
          0 & 0 & \frac{p{+}\Pi{+}r^2\pi^{\phi\phi}}{r^2} & 0\\
          0 & 0 & 0 & \frac{p{+}\Pi{+}\tau^2\pi^{\eta\eta}}{\tau^2} 
      \end{array}
\right).
\end{equation}

The non-vanishing components of the shear tensor 
$\sigma^{mn}{\,=\,}\nabla^{\left\langle m\right.} u^{\left.n\right\rangle}$
are found (after some algebra which properly takes into account the
Christoffel symbol contributions to the covariant derivatives) to be
\begin{subequations}
\label{A11}
\begin{eqnarray}
\label{A11a}
\sigma^{\eta\eta} &=& \frac{1}{\tau^2} \left(\frac{\theta}{3}
                     -\frac{\gamma_r}{\tau}\right),
\\
\label{A11b}
\sigma^{\phi\phi} &=& \frac{1}{r^2} \left(\frac{\theta}{3}
                     -\frac{\gamma_r v_r}{r}\right),
\\
\label{A11c}
  \sigma^{rr} &=& -\gamma_r^2 
  \left(\frac{2\theta}{3}-\frac{\gamma_r}{\tau}-\frac{\gamma_r v_r}{r}\right)
\nonumber\\
  &=& \gamma_r^2 \left(\frac{\theta}{3}-\partial_\tau\gamma_r
                                     -\partial_r(\gamma_r v_r)\right),
\\
\label{A11d}
  \sigma^{\tau\tau} &=& (1-\gamma_r^2) 
  \left(\frac{2\theta}{3}-\frac{\gamma_r}{\tau}-\frac{\gamma_r v_r}{r}\right),
\\
\label{A11e}
\sigma^{\tau r} &=& v_r \sigma^{rr}.
\end{eqnarray}
\end{subequations}
Note that (\ref{A11b}) differs from the corresponding result (44) in 
\cite{MR04} by a metric factor $r^2$, and that the last expression in 
(\ref{A11c}) has an extra factor $\gamma_r^2$ when compared to Eq.~(43) 
in \cite{MR04}.

\subsection{Systems without azimuthal symmetry}
\label{appb}
Without azimuthal symmetry, there is no advantage in using transverse
polar coordinates, and it is simpler to use Cartesian coordinates
in the transverse plane: $x^m{\,=\,}(\tau,x,y,\eta)$. The global frame
and the local fluid rest frame are now related by a Lorentz boost
with velocity $\bm{v}_\perp{\,=\,}v_x(x,y,\tau)\,\bm{e}_x + v_y(x,y,\tau)\,
\bm{e}_y=(v_x,v_y)$, described by the Lorentz transformation matrix
\begin{equation}
\label{A15}
{\Lambda^m}_n(\bm{v}_\perp){\,=}\left(
   \begin{array}{cccc}
          \gamma_\perp     & \gamma_\perp v_x & \gamma_\perp v_y & 0\\ 
          \gamma_\perp v_x & 1 + (\gamma_\perp{-}1)\frac{v_x^2}{v_\perp^2}
                           & (\gamma_\perp{-}1)\frac{v_x v_y}{v_\perp^2} & 0\\ 
          \gamma_\perp v_y & (\gamma_\perp{-}1)\frac{v_x v_y}{v_\perp^2} & 
          1 + (\gamma_\perp{-}1)\frac{v_y^2}{v_\perp^2} & 0\\
                0      &    0      & 0 & 1
      \end{array}
\right)\!.
\end{equation}
The projector transverse to the flow vector $u^m$ takes the form
\begin{equation}
\label{A16}
{\Delta^m}_n = \gamma_\perp^2 
\left(\begin{array}{cccc}
            -v_\perp^2 & v_x        & v_y       & 0\\ 
            -v_x       & 1{-}v_y^2  & v_x v_y   & 0\\ 
            -v_y       & v_x v_y    & 1{-}v_x^2 & 0\\
                0      &    0       & 0         & 1
      \end{array}
\right).
\end{equation}
In the local rest frame the energy-momentum tensor reads
\begin{equation}
\label{A17}
\hat T^{mn} = 
\left(\begin{array}{cccc}
          e & 0 & 0 & 0\\ 
          0 & p{+}\Pi & 0 & 0\\
          0 & 0 & p{+}\Pi & 0\\
          0 & 0 & 0 & \frac{p{+}\Pi}{\tau^2} 
      \end{array}
\right) + 
\left(\begin{array}{cccc}
          0 & 0 & 0 & 0\\ 
          0 & \hat\pi^{xx} & \hat\pi^{xy} & 0\\
          0 & \hat\pi^{xy} & \hat\pi^{yy} & 0\\
          0 & 0 & 0 & \hat\pi^{\eta\eta} 
      \end{array}
\right).
\end{equation}
The shear pressure tensor $\hat\pi^{mn}$ is no longer diagonal. It may 
be useful to have expressions for the shear pressure tensor components
$\hat\pi^{mn}$ in the local rest frame in terms of the hydrodynamic
solution for the energy-momentum tensor $T^{mn}$ in the global frame.
To this end we follow Ref.~\cite{MR04} and introduce the mutually
orthogonal rest frame 4-vectors
\begin{eqnarray}
\label{A18}
&&\hat u^m = (1,0,0,0), \quad
  \hat i^m = (0,1,0,0), 
\nonumber\\
&&\hat j^m = (0,0,1,0), \quad
  \hat h^m = (0,0,0,1/\tau),
\end{eqnarray}
with $\hat u{\cdot}\hat u{\,=\,}1$, $\hat i{\cdot}\hat i{\,=\,}\hat 
j{\cdot}\hat j{\,=\,}\hat h{\cdot}\hat h{\,=\,}-1$, such that  
\begin{subequations}
\label{A19}
\begin{eqnarray}
\label{A19a}
\hat \pi^{xy} &=& {\hat i}_m \hat T^{mn} {\hat j}_n\equiv (i{\cdot}T{\cdot}j),
\\
\label{A19b}
\hat p+\Pi &=& \frac{1}{3} \Bigl[(i{\cdot}T{\cdot}i)+(j{\cdot}T{\cdot}j)
                                +(h{\cdot}T{\cdot}h)\Bigr],
\\
\label{A19c}
\hat \pi^{xx} &=& (i{\cdot}T{\cdot}j) - p -\Pi 
\nonumber\\
  &=& \frac{2}{3}(i{\cdot}T{\cdot}i)
  - \frac{1}{3}\Bigl((j{\cdot}T{\cdot}j)+(h{\cdot}T{\cdot}h)\Bigr),
\\
\label{A19d}
\hat \pi^{yy} &=& \frac{2}{3}(j{\cdot}T{\cdot}j)
  - \frac{1}{3}\Bigl((i{\cdot}T{\cdot}i)+(h{\cdot}T{\cdot}h)\Bigr),
\\
\label{A19e}
\tau^2 \hat \pi^{\eta\eta} &=& \frac{2}{3}(h{\cdot}T{\cdot}h)
  - \frac{1}{3}\Bigl((i{\cdot}T{\cdot}i)+(j{\cdot}T{\cdot}j)\Bigr).
\qquad
\end{eqnarray}
\end{subequations}
The right hand sides are Lorentz invariant expressions and thus can be 
evaluated in any reference frame. The vectors $u^m,\,i^m,\,j^m,\,h^m$
in the global frame are obtained by applying the Lorentz boost 
(\ref{A15}) to the rest frame vectors (\ref{A18}):
\begin{subequations}
\label{A24}
\begin{eqnarray}
\label{A24a}
&&u^m = \gamma_\perp(1,v_x,v_y,0),
\\
\label{A24b}
&&i^m = \Bigl(\gamma_\perp v_x,
              1{+}(\gamma_\perp{-}1)\frac{v_x^2}{v_\perp^2},
              (\gamma_\perp{-}1)\frac{v_x v_y}{v_\perp^2},
              0\Bigr), \qquad\quad
\\
\label{A24c}
&&j^m = \Bigl(\gamma_\perp v_y,
              (\gamma_\perp{-}1)\frac{v_x v_y}{v_\perp^2},
              1{+}(\gamma_\perp{-}1)\frac{v_y^2}{v_\perp^2},
              0\Bigr), \qquad\quad
\\
\label{A24d}
&&h^m = \left(0,0,0,\frac{1}{\tau}\right).
\end{eqnarray}
\end{subequations}
From (\ref{A24d}) it follows immediately that the $\eta\eta$ component
of $T^{mn}$ is identical in the global and fluid rest frames, i.e. 
that $\pi^{\eta\eta}{\,=\,}\hat\pi^{\eta\eta}$. Longitudinal boost
invariance implies that
\begin{equation}
\label{A28} 
\sigma^{\eta\tau}=\sigma^{\eta x}=\sigma^{\eta y}=0, \qquad
\pi^{\eta\tau}=\pi^{\eta x}=\pi^{\eta y}=0.
\end{equation}
The constraints from tracelessness,
\begin{equation}
\label{A29}
\pi^{\tau\tau}=\pi^{xx}+\pi^{yy} + \tau^2 \pi^{\eta\eta},
\end{equation}
and orthogonality to $u^m$,
\begin{subequations}
\label{A30}
\begin{eqnarray} 
\label{A30a}
&&\pi^{x\tau} = v_x \pi^{xx} + v_y \pi^{xy},
\\
\label{A30b}
&&\pi^{y\tau} = v_x \pi^{xy} + v_y \pi^{yy},
\\
&&\label{A30c}
\pi^{\tau\tau} = v_x \pi^{\tau x} + v_y \pi^{\tau y},
\end{eqnarray}
\end{subequations}
then leave us with three independent components of $\pi^{mn}$. Which
should we select? From the above it is obvious that $\pi^{\eta\eta}$
should be one of them. For the other two there are two different
possibilities, following from two different chains of reasoning, as
explained in Sections~\ref{sec5b1} and \ref{sec5b2}.

\begin{widetext}
Before giving the explicit expressions for the shear tensor needed 
in each case, let us first generalize the relations (\ref{A7}), 
which are needed to bring the hydrodynamic equations into standard form:
\begin{subequations}
\label{A31}
\begin{eqnarray}
\label{A31a}
&& T^{\tau\tau}=(e{+}p{+}\Pi)\gamma_\perp^2 - (p{+}\Pi) + \pi^{\tau\tau},
\\
\label{A31b}
&& T^{\tau x}=(e{+}p{+}\Pi)\gamma_\perp^2 v_x + \pi^{\tau x} 
             = (T^{\tau\tau}{+}p{+}\Pi{-}\pi^{\tau\tau})v_x + \pi^{\tau x},
\\
\label{A31c}
&& T^{\tau y}=(e{+}p{+}\Pi)\gamma_\perp^2 v_y + \pi^{\tau y} 
             = (T^{\tau\tau}{+}p{+}\Pi{-}\pi^{\tau\tau})v_y + \pi^{\tau y},
\\
\label{A31d}
&& T^{xx}=(e{+}p{+}\Pi)\gamma_\perp^2 v_x^2 + p + \Pi + \pi^{xx} 
             = (T^{\tau x}{-}\pi^{\tau x})v_x + p + \Pi + \pi^{xx},
\\
\label{A31e}
&& T^{xy}=(e{+}p{+}\Pi)\gamma_\perp^2 v_x v_y + \pi^{xy} 
             = (T^{\tau x}{-}\pi^{\tau x})v_y + \pi^{xy}
             = (T^{\tau y}{-}\pi^{\tau y})v_x + \pi^{xy},
\\
\label{A31f}
&& T^{yy}=(e{+}p{+}\Pi)\gamma_\perp^2 v_y^2 + p + \Pi + \pi^{yy} 
             = (T^{\tau y}{-}\pi^{\tau y})v_y + p + \Pi + \pi^{yy},
\end{eqnarray}
\end{subequations}

If we choose $\pi^{\tau\tau}$, $\pi^{\eta\eta}$ and 
$\Delta{\,=\,}\pi^{xx}{-}\pi^{yy}$ as independent dynamical variables,
we can solve the constraints (\ref{A29}), (\ref{A30}) to gives the 
following expressions for the dependent components of $\pi^{mn}$:
\begin{subequations}
\label{A32}
\begin{eqnarray}
\label{A32a} 
&&2v_x\pi^{\tau x} = \pi^{\tau\tau}\left(1+\frac{v_x^2{-}v_y^2}{2}\right)
  - \tau^2 \pi^{\eta\eta}\frac{v_x^2{-}v_y^2}{2}+\frac{v_\perp^2}{2}\Delta,
\\
\label{A32b} 
&&2v_y\pi^{\tau y} = \pi^{\tau\tau}\left(1-\frac{v_x^2{-}v_y^2}{2}\right)
  + \tau^2 \pi^{\eta\eta}\frac{v_x^2{-}v_y^2}{2}-\frac{v_\perp^2}{2}\Delta,
\\
\label{A32c} 
&&2v_xv_y\pi^{xy} = \pi^{\tau\tau}\left(1-\frac{v_\perp^2}{2}\right)
  + \tau^2 \pi^{\eta\eta}\frac{v_\perp^2}{2}-\frac{v_x^2{-}v_y^2}{2}\Delta,
\\
\label{A32d} 
&&\pi^{xx} = \frac{1}{2}
             \Bigl(\pi^{\tau\tau}-\tau^2 \pi^{\eta\eta}+\Delta\Bigr),
\\
\label{A32e} 
&&\pi^{yy} = \frac{1}{2}
             \Bigl(\pi^{\tau\tau}-\tau^2 \pi^{\eta\eta}-\Delta\Bigr).
\end{eqnarray}
\end{subequations}
In this approach we need the following components of the stress
tensor $\sigma^{mn}$ as source terms for the kinetic transport equations
for $\pi^{\tau\tau}$, $\pi^{\eta\eta}$, and $\Delta$:
\begin{subequations}
\label{A37}
\begin{eqnarray}
\label{A37a} 
&&\sigma^{\tau\tau} = \frac{\theta}{3}(\gamma_\perp^2-1) 
  + \partial_\tau\gamma_\perp-\frac{1}{2}D(\gamma_\perp^2),
\\
\label{A37b} 
&&\sigma^{\eta\eta} = \frac{1}{\tau^2}
  \left(\frac{\theta}{3}-\frac{\gamma_\perp}{\tau}\right),
\\
\label{A37c} 
&&\sigma^{\Delta} = \sigma^{xx}{-}\sigma^{yy} = 
  \frac{\theta}{3}\Bigl(2+\gamma_\perp^2(v_x^2{-}v_y^2)\Bigr)
  +\partial_y(\gamma_\perp v_y)-\partial_x(\gamma_\perp v_x)
  -\frac{1}{2}D\Bigl(\gamma_\perp^2(v_x^2{-}v_y^2)\Bigr).
\end{eqnarray}
\end{subequations}

If instead of $\Delta$ we keep $\pi^{\tau x}$, $\pi^{\tau y}$ as
dynamical variables (as explained in Section~\ref{sec5b2}), we don't 
need Eqs.~(\ref{A32a}) and (\ref{A32b}), but we still must express 
$\pi^{xy}$, $\pi^{xx}$, and $\pi^{yy}$ (which appear as sources
on the right hand sides of Eqs.~(\ref{eq59b}) and (\ref{eq59c})) in terms
of those components for which we solve kinetic transport equations.
To this end we rewrite the constraints (\ref{A29}), (\ref{A30}) as
\begin{subequations}
\label{A40}
\begin{eqnarray}
\label{A40a} 
&&\pi^{xx} = \frac{1}{v_\perp^2}
  \Bigl(v_y^2(\pi^{\tau\tau}{-}\tau^2\pi^{\eta\eta}) 
        + v_x\pi^{\tau x} - v_y \pi^{\tau y}\Bigr),
\\
\label{A40b} 
&&\pi^{yy} = \frac{1}{v_\perp^2}
  \Bigl(v_x^2(\pi^{\tau\tau}{-}\tau^2\pi^{\eta\eta}) 
        + v_y\pi^{\tau y} - v_x \pi^{\tau x}\Bigr),
\\
\label{A40c} 
&&\pi^{xy} = -\frac{1}{v_\perp^2}
  \Bigl(v_x v_y(\pi^{\tau\tau}{-}\tau^2\pi^{\eta\eta}) 
        - v_x\pi^{\tau y} - v_y \pi^{\tau x}\Bigr).
\end{eqnarray}
\end{subequations}
The shear tensor components required on the right hand sides of 
Eqs.~(\ref{eq73}) are
\begin{subequations}
\label{A43}
\begin{eqnarray}
\label{A43a} 
&&\sigma^{\tau x} = - \frac{1}{2}\partial_x\gamma_\perp 
                    + \frac{1}{2}\partial_\tau(\gamma_\perp v_x)
                    - \frac{1}{2} D(\gamma_\perp^2 v_x)
                    + \frac{\theta}{3}\,\gamma_\perp^2v_x,
\\
\label{A43b} 
&&\sigma^{\tau x} = - \frac{1}{2}\partial_y\gamma_\perp 
                    + \frac{1}{2}\partial_\tau(\gamma_\perp v_y)
                    - \frac{1}{2} D(\gamma_\perp^2 v_y)
                    + \frac{\theta}{3}\,\gamma_\perp^2v_y.
\end{eqnarray}
\end{subequations}
\end{widetext}


\end{document}